\documentclass[12pt]{article}
\usepackage{a4}
\usepackage{epsfig}
\usepackage{color}
\begin{document}

\bibliographystyle{unsrt}

\begin{flushright}
IFT 2002/21
\end{flushright}
\begin{center}
\vspace{0.6cm}
\begin{Large}
Contributions  due to 
the longitudinal virtual photon  
in the semi-inclusive $ep$ collision at HERA\footnote{
Presented at X International Workshop on Deep Inelastic Scattering
(DIS 2002),\\ Cracow, 30 April - 4 May 2002}\\
\vspace{1cm}
Urszula Jezuita-D\c{a}browska\\
\end{Large}
\vspace{0.4cm}
{\sl Institute of Theoretical Physics, Warsaw University}\\
{\sl ul. Ho\.za 69, 00-681 Warsaw, Poland}\\
\vspace{0.5cm}
\end{center}



\begin{center}
\bf Abstract
\end{center}
The importance of  contributions due to the longitudinally polarised 
virtual photon, $d \sigma_L$
and the interference term $d \tau_{LT}$,
in the unpolarised $e p$ collisions is discussed \cite{jdk}.
The numerical calculations 
for the  Compton process
$e p \rightarrow e \gamma X$ 
at the HERA collider were performed
in the Born approximation.
The various distributions
in the $CM_{ep}$ and Breit frames are presented.
These cross sections
are dominated by
the transversely polarised  intermediate photon,
even for large $Q^2$.

\section{Introduction}\label{sec:intr}
\vspace{-0.2cm}
In cross sections for
semi-inclusive $ep$ processes and
collisions with  two intermediate photon, 
the terms 
coming from the interference between $\gamma_L^{\ast}$ and $\gamma_T^{\ast}$
or between two different transverse states of $\gamma^{\ast}$
can appear \cite{bv}.
The detailed studies of various contributions 
for the process $e^+ e^- \rightarrow e^+ e^- \mu^+ \mu^-$
performed for the kinematical range of
the PLUTO and LEP experiments\cite{ai} 
show the importance of interference terms.

Here we study
the longitudinal-transverse interference term ($d\tau_{LT}$)
and contributions due to exchange of 
$\gamma^{\ast}_L$ ($d\sigma_L$) and  $\gamma^{\ast}_T$ ($d\sigma_T$)
in the unpolarised semi-inclusive
$ep$ collisions \cite{jdk}.
Assuming one-photon exchange
we factorise the  cross-section
onto the photon emission by the electron 
and the $\gamma^{\ast} p$ collision
in a way  independent on the reference frame.
For this purpose  we
use the propagator decomposition method 
and  explicit forms of
all polarisation vectors
of the virtual photon ($q^2 < 0$).

\vspace{-0.3cm}
\section{Factorisation formulae for unpolarised $ep$ collisions}\label{sec:fac}
\vspace{-0.2cm}
The cross section for an unpolarised $l N \rightarrow l X$ process,
for example DIS $e p \rightarrow e X$,
can be factorised
onto the leptonic and hadronic tensors,
$d\sigma \; \sim \; L^{\mu\nu} W_{\mu\nu}$.
Further on the differential cross section can be decomposed on the parts 
related to 
the subprocesses $\gamma^{\ast}_T N \rightarrow X$ and 
$\gamma^{\ast}_L N \rightarrow X$,
respectively:
\begin{equation}
d\sigma^{e p \rightarrow e X} \; = \; 
\Gamma_T \; \sigma^{\gamma^{\ast} p \rightarrow X}_T \; + \; 
\Gamma_L \; \sigma^{\gamma^{\ast} p \rightarrow X}_L \; .
\label{eq:sub21a}
\end{equation}
The above factorisation and separation formula 
can be obtained in  various  ways. One of them uses the known hadronic
tensor and explicit form of the scalar polarisation vector\cite{ha}.
Another way 
is the propagator decomposition method\cite{kr}
in which the cross section is written as follows
\begin{equation}
d \sigma^{e p \rightarrow e X} \; \sim \; L_e^{\alpha\beta} \; \frac{g_{\alpha\mu}}{q^2} \; \frac{g_{\nu\beta}}{q^2} \; W_p^{\mu\nu} \; .
\label{eq:sub21b}
\end{equation}
Afterwards one decomposes the propagator of the exchanged photon
using
the completeness relation, what leads directly to Eq.~(\ref{eq:sub21b}).
This method is especially useful in analysing of 
the semi-inclusive processes.

\begin{figure}[h]
\vspace{-24.0cm}
\hspace{1cm}
\epsfig{file=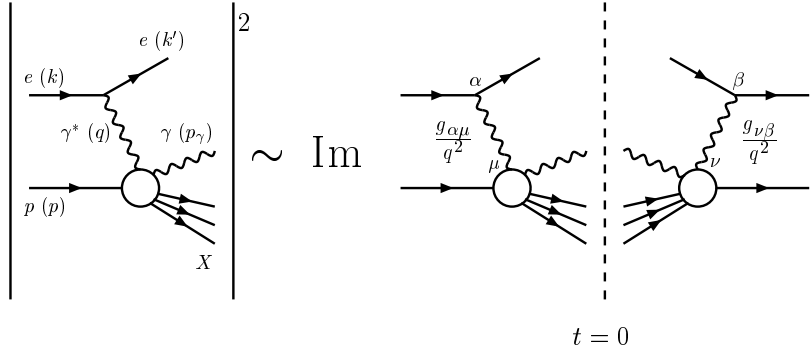}
\vspace{-1.2cm}
\caption{The optical theorem for
the Compton process $e p \rightarrow e \gamma X$. \label{fig:diagram2}}
\vspace{-0.5cm}
\end{figure}
\vspace{0.5cm}
In case of the semi-inclusive process 
one additional particle in the final state is produced.
For example for the Compton process
$e p \rightarrow e \gamma X$ (Fig.~\ref{fig:diagram2})
the differential cross section  can be decomposed as follows:
\begin{equation}
d\sigma^{e p \rightarrow e \gamma X}  \; = \; 
d \sigma_{T}^{e p \rightarrow e \gamma X}
\; + \; 
d \sigma_{L}^{e p \rightarrow e \gamma X}
\; + \; d \tau_{TT}^{e p \rightarrow e \gamma X}  \; + \; 
d \tau_{LT}^{e p \rightarrow e \gamma X} \; .
\end{equation}
In the above formula two additional contributions, 
$d \tau_{LT}$ and $d \tau_{TT}$, appear.
They are related to the interference between 
$\gamma_L^{\ast}$ and $\gamma_T^{\ast}$, and
between two 
different transverse polarisation
states of the $\gamma^{\ast}$, respectively.\\

\begin{figure}[h]
\vspace{-14.5cm}
\hspace{1cm}
\epsfig{file=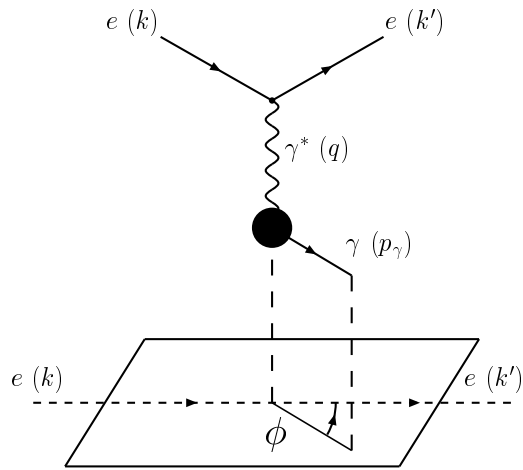}
\vspace{-9.5cm}
\caption{The azimuthal angle $\phi$ for the process  $e p \rightarrow e \gamma X$ in the Breit frame. \label{fig:diagram3}}
\end{figure}
\vspace{0.5cm}
In studies of the interference terms in the semi-inclusive processes 
$e p \rightarrow e \gamma X$ 
the azimuthal angle $\phi$ distribution is especially useful. 
The angle $\phi$ is defined as
the difference of the azimuthal angle of the final
electron  and of the final photon:
$\phi \; = \; \phi_e \; - \; \phi_{\gamma}$.

In  the Breit frame 
$\phi$ is the angle between the electron scattering plane and 
plane fixed by the momenta of the exchanged $\gamma^{\ast}$
and final photon $\gamma$.
In this  reference frame  
$d \sigma / d \phi$ is linear in 
$\cos \phi$, $\cos 2\phi$, $\sin \phi$ and $\sin 2\phi$.
For calculations in the Born approximation
the terms containing $\sin \phi$ and $\sin 2\phi$  vanish
as a consequence of time-reversal invariance,
so the azimuthal distribution for the
Compton process reduces 
to the following form
\cite{Brown:1971}: 
\begin{equation}
\frac{d \sigma^{e p \rightarrow e \gamma X}}{d \phi} \; = \;
\sigma_0 \; + \; \sigma_1 \; \cos \phi \; + \; \sigma_2 \; \cos 2\phi \; .
\label{eq:mkj:t7}
\end{equation}
The coefficients $\sigma_0$, $\sigma_1$ and $\sigma_2$
are related to 
$d \sigma_T / d \phi$, $d \sigma_L / d \phi$, $d \tau_{LT} / d \phi$ and  
$d \tau_{TT} / d \phi$. 
The third term arises from the interference between 
two different transverse polarisation states of the  exchanged photon 
($\sigma_2 \cos 2\phi = d \tau_{TT}/d \phi$).
The longitudinal-transverse interference 
gives rise to the second term ($\sigma_1  \cos \phi = d \tau_{LT} / d \phi$).
The $\sigma_0$ consists of the sum of
the cross sections with the intermediate  
$\gamma_L^{\ast}$ and $\gamma_T^{\ast}$
($\sigma_0 =  d \sigma_L / d \phi + d \sigma_{T} / d \phi$).
Therefore the $\phi$ distribution in the Breit frame
is an excellent tool to identify and study interference terms.

\vspace{-0.3cm}
\section{Numerical results for  Compton process $e p \rightarrow e \gamma X$}\label{subsec:num}

\begin{figure}[h]
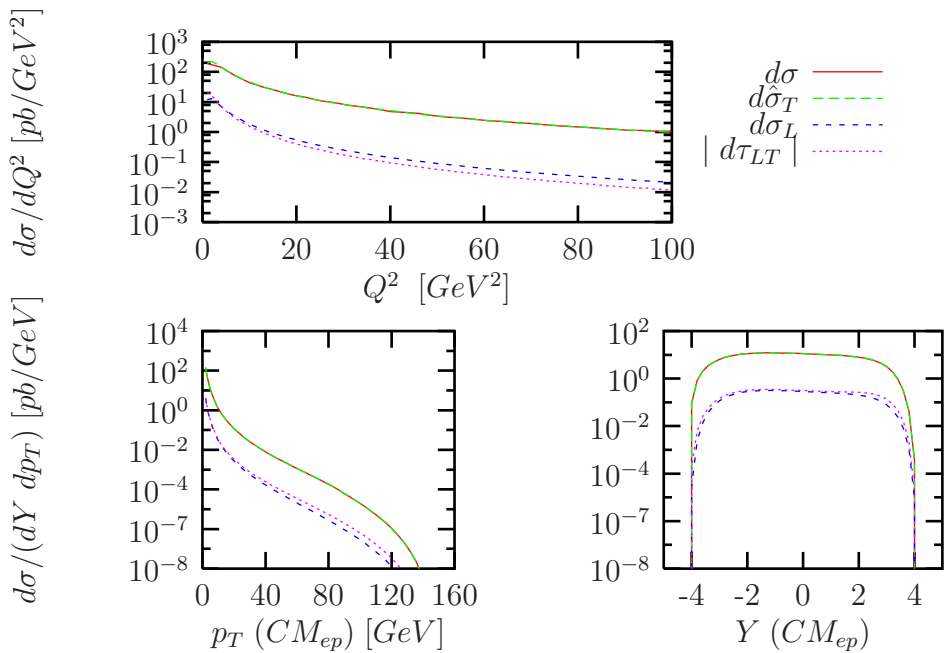

{\input{SAepQ.tex} \\
\input{SAepPTY.tex} \input{SAepYPT.tex}}
\caption{Contributions to $d \sigma / d Q^2$ (at the top)
and to $d \sigma / (dp_T dY)$ (below)
as a functions of $p_T$ with $Y = 0$ 
(on left)
or $Y$ with $p_T = 5$ GeV (on right), in $CM_{ep}.$ 
\label{fig:wykres1}}
\end{figure}

We calculate various contributions to the cross sections
for the unpolarised Compton process $e p \rightarrow e \gamma X$
in both the $CM_{ep}$ and Breit frames 
for the HERA energy $\sqrt{S_{ep}} \; = \; 300$ GeV.
We consider the emission of the $\gamma$ from the hadronic vertex
at the Born level 
(i.e.  
the $\gamma^{\ast} q \rightarrow \gamma q$ subprocess only)\footnote{
The cross section for the 
Bethe-Heitler process, i.e. production of 
the $\gamma$  from the electron line, can be neglected
for the photon's rapidity range  $Y (CM_{ep}) < 0$
\cite{ke}.}.
For the proton we have used the CTEQ5L parton parametrization 
\cite{li} with $N_f = 4$ 
and the hard scale equals to $p_T$.

The cross section $d \sigma / d Q^2$, (Fig.~\ref{fig:wykres1}, top)
is strongly dominated by 
contribution due to the transversely polarised $\gamma^{\ast}$,
even for large values of virtuality $Q^2$.
Also the cross sections  $d \sigma / (dp_T dY)$
(Fig.~\ref{fig:wykres1}, bottom), as a function of
$p_T$ or rapidity $Y$, are very well described by the $\gamma^{\ast}_T$ 
cross section only .
Both contributions coming from the $\gamma_L^{\ast}$,
$d \sigma_{L}$ and $d \tau_{LT}$,
are below $10 \%$,
moreover due to   opposite signs they almost cancel each other.\\
\begin{figure}[h]
\hspace{1.5cm}
\input{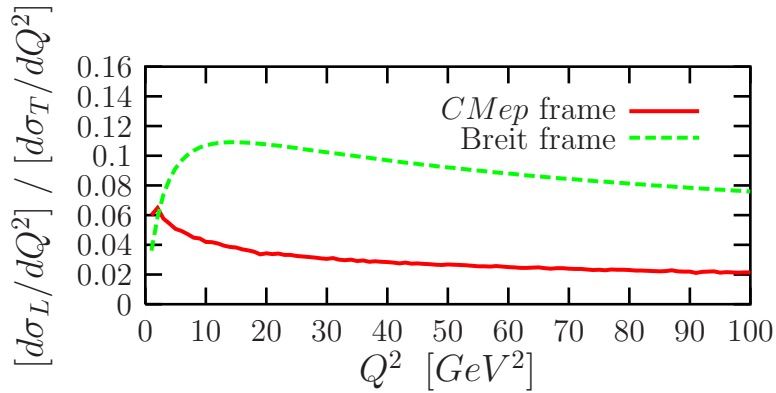}
\caption{The ratio 
$[d\sigma_L/dQ^2] \; / \; [d\sigma_T/dQ^2]$ 
as a function of $Q^2$, in the $CM_{ep}$ frame (solid line) 
and in the Breit frame (dashed line).}
\label{fig:wykres2}
\end{figure}

The ratio 
$[d\sigma_L/dQ^2] \; / \; [d\sigma_T/dQ^2]$ 
(Fig.~\ref{fig:wykres2})
shows interesting $Q^2$ dependence
in two reference frames ($CM_{ep}$ 
and Breit frame).
We see that domination of
the cross sections by
$\gamma_T^{\ast}$
is stronger in the $CM_{ep}$ frame 
in which $d \sigma_L$ and $d \tau_{LT}$  
almost cancel each other.\\

\begin{figure}[h]
\vspace{-5.0cm}
\hspace{-1cm}
\epsfxsize=50pc 
\epsfbox{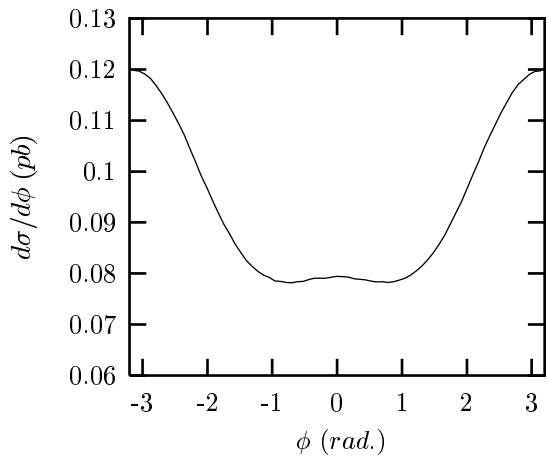}
\vspace{-19.0cm}
\caption{The $d \sigma / d \phi$ in the Breit frame. \label{fig:adphi}}
\end{figure}

For  the azimuthal angle distribution
in  Breit frame
the relatively large sensitivity 
to the interference term $d \tau_{LT}$
is found (Fig.~\ref{fig:adphi}), while the interference
between two different transverse polarisation states of $\gamma$
is invisible.

\vspace{-0.6cm}
\section{Conclusions}\label{subsec:con}
\vspace{-0.2cm}
Our analysis show that
the cross section for the  Compton process
(the Born level) in $CM_{ep}$ is strongly dominated by $\gamma^{\ast}_T$.
If the contributions due to  $\gamma_{L}^{\ast}$
are included then
interference terms need to be included 
in a consistent analysis because they both are similar 
in size but opposite in sign.

The studies of the azimuthal angle dependence,
$d\sigma^{ep \rightarrow e \gamma X} / d \phi$,
in the Breit frame give access to the 
longitudinal-transverse interference term.\\

I would like to acknowledge Maria Krawczyk for fruitful discussions
and for reading manuscript.\\

\end{document}